\documentclass[a4paper,11pt]{article}
\usepackage{pos}

\usepackage{array} % array
\usepackage{bm} % bold math
\usepackage{hyperref}
\usepackage{multirow}
\usepackage{xcolor}
\usepackage{subfigure}
\usepackage{adjustbox}
\usepackage{enumitem}
\usepackage{caption}
\usepackage{anyfontsize}
\usepackage{color}
\graphicspath{{./figs/}}

\allowdisplaybreaks

\newcommand{\dslash}[1]{\text{$\not \!\! #1$}}
\newcommand{\Tr}{\mathrm{Tr}}

\newcolumntype{\.}{>{\global\let\currentrowstyle\relax}}

\newcolumntype{^}{>{\currentrowstyle}}
\newcommand{\rowstyle}[1]{\gdef\currentrowstyle{#1}%
    #1\ignorespaces
}

\newcommand{\red}[1]{\textcolor{red}{#1}} 
\newcommand{\blue}[1]{\textcolor{blue}{#1}}

\title{Machine Learning Estimation on the Trace of Inverse Dirac Operator 
using the Gradient Boosting Decision Tree Regression}
\ShortTitle{Trace estimation with gradient boosting decision tree regression}

\author*[a]{Benjamin J. Choi}
\author[a]{Hiroshi Ohno}
\author[b]{Takayuki Sumimoto}
\author[c]{Akio Tomiya}

\affiliation[a]{Center for Computational Sciences, 
University of Tsukuba,\\
1-1-1 Tennodai, Tsukuba, Ibaraki 305-8577, Japan}

\affiliation[b]{FLECT Co.,~Ltd., 
1-1-1 Shibaura, Minato-ku, Tokyo 105-0023, Japan}

\affiliation[c]{Department of Information and Mathematical Sciences,
Tokyo Woman’s Christian University,\\
2-6-1 Zempukuji, Suginami-ku, Tokyo 167-8585, Japan}

\emailAdd{benchoi@ccs.tsukuba.ac.jp}
\emailAdd{hohno@ccs.tsukuba.ac.jp}
\emailAdd{akio@yukawa.kyoto-u.ac.jp}

\abstract{We present our preliminary results on the machine learning estimation
of $\Tr \, M^{-n}$ from other observables with the gradient boosting decision
tree regression, where $M$ is the Dirac operator. 
Ordinarily, $\Tr \, M^{-n}$ is obtained by linear CG solver for stochastic
sources which needs considerable computational cost. 
Hence, we explore the possibility of cost reduction on the trace estimation by
the adoption of gradient boosting decision tree algorithm.
We also discuss effects of bias and its correction.}

\FullConference{The 41st International Symposium on Lattice Field Theory
(LATTICE2024) \\
28 July - 3 August 2024 \\
Liverpool, UK \\}

\begin{document}
%--------------------------------------------

%--------------------------------------------
\maketitle
%--------------------------------------------

%--------------------------------------------
\section{Introduction}
\label{sec:intro}
%--------------------------------------------

For Lattice QCD calculations on observables such as cumulants of the chiral
order parameter, the trace of operators ($\Tr \, O$) is often necessary where
$O$ represents an observable such as $O = M^{-1}$, inverse Dirac operator.
Here, the trace of operator is often estimated with a stochastic source method
\cite{Dong:1993pk}.
Since $M= \left( \dslash{D} + m_0 \right)$, the Dirac operator, is a large
sparse matrix on the lattice, its inverse matrix obtained by a linear CG solver
requires a considerable computational cost.

In this paper, we present our preliminary result on machine learning estimation
of $\Tr \, M^{-n}$ from other observables such as $\Tr \, M^{-m}$ ($m < n$).
As input observables, we also use plaquette and Polyakov loop obtained
when generating gauge configurations with the hybrid Monte Carlo (HMC)
algorithm \cite{Duane:1985hz,Duane:1986iw,Duane:1987de}.
Here, we use the gradient boosting decision tree regression method
\cite{10.1214/aos/1013203451}, based on the methodology of Yoon \emph{et
al.}~\cite{Yoon:2018krb}.
Note that recently, a machine learning mapping between $\Tr \, M^{-1}$ at
different quark masses and gradient flow times was studied in the similar way
\cite{Kim:2024rpd}.

For this preliminary analysis, we use the data originally produced for
Ref.~\cite{Ohno:2018gcx} where Oakforest-PACS system \cite{Boku:2017urp} and
BQCD program \cite{Nakamura:2010qh} are used.
In this data, we use $N_f=4$ Wilson clover action \cite{Sheikholeslami:1985ij}
and Iwasaki gauge action \cite{Iwasaki:1985we,Iwasaki:1983iya} as explained in
Ref.~\cite{Ohno:2018gcx}.
The data used in this work are given in Table~\ref{tab:ensemble-1}.
We choose 3 datasets which have the same lattice size of $16^3 \times 4$, the
lattice coupling constant $\beta=1.60$, the clover coefficient $c_{\text{SW}} =
2.065$ and the number of gauge configurations $N_{\text{conf}} = 5500$ but
different $\kappa$ values.
Especially, at the \texttt{ID-1} in Table~\ref{tab:ensemble-1}, a first phase
transition is observed.
\begin{table}[tb]
    \vspace{-1.0em}
    \renewcommand{\arraystretch}{1.2}
    \centering
    \begin{adjustbox}{max width=\textwidth}
    \begin{tabular}{@{\quad}\.c@{\quad}|@{\qquad}^c@{\qquad}|@{\quad}^c@{\quad}|@{\qquad}^c@{\qquad}^c@{\qquad}|@{\quad}^c@{\quad}}
    \hline
    \hline
    \texttt{ID} & $L^3 \times T$  & $\beta$ & $\kappa$ & $c_{\textrm{SW}}$ & $N_{\textrm{conf}}$ \\
    \hline
    \texttt{0}  & $16^3 \times 4$ & 1.60 & 0.13575 & 2.065 & 5500 \\
    %\texttt{1}  & $16^3 \times 4$ & 1.60 & 0.13577 & 2.065 & 5500 \\
    \rowstyle{\bfseries} \texttt{1}  & $\mathbf{16^3 \times 4}$ & 1.60 & 0.13580
    & 2.065 & 5500 \\
    %\texttt{3}  & $16^3 \times 4$ & 1.60 & 0.13582 & 2.065 & 5500 \\
    \texttt{2}  & $16^3 \times 4$ & 1.60 & 0.13585 & 2.065 & 5500 \\
    \hline
    \hline
    \end{tabular}
    \end{adjustbox}
    \caption{Data used in this paper.
    These data are originally produced for Ref.~\cite{Ohno:2018gcx}.
    Note that the first phase transition is observed in the \texttt{ID-1} (the
    row with bold text).}
    \label{tab:ensemble-1}
\end{table}
%\rowstyle{\color{red}}

%--------------------------------------------
\section{Analysis detail}
\label{sec:anly}
%--------------------------------------------

\begin{table}[tb]
    \vspace{-1.5em}
    \renewcommand{\arraystretch}{1.2}
    \centering
    \begin{adjustbox}{max width=\textwidth}
    \begin{tabular}{@{\quad}\.r@{\quad}|@{\quad}^l@{\quad}}
    \hline
    \hline
    Symbol & Description \\
    \hline
    $X$, $Y$ & observables used as input ($X$) and output ($Y$) (\emph{e.g.},~$X = \Tr \,
    M^{-1}$, $Y = \Tr \, M^{-2}$) \\
    $S^Z$ & the total dataset of $Z=X,Y$ where $S^Z = S^Z_{\text{LB}} \cup
    S^Z_{\text{UL}}$ \\
    $S^Z_{\text{LB}}$ & the labeled set of $Z=X,Y$ where $S^Z_{\text{LB}} =
    S^Z_{\text{TR}} \cup S^Z_{\text{BC}}$ \\
    $S^Z_{\text{TR}}$ & the training set of $Z=X,Y$ \\
    $S^Z_{\text{BC}}$ & the bias correction set of $Z=X,Y$ \\
    $S^Z_{\text{UL}}$ & the unlabeled set of $Z=X,Y$ \\
    $N$ & the number of elements of $S^Z$ where $N = \left|S^{X}\right| =
    \left|S^{Y}\right|$ \\
    $N_{\text{LB}}$ & the number of elements of $S^Z_{\text{LB}}$ where
    $N_{\text{LB}} = \left|S^{X}_{\text{LB}}\right| =
    \left|S^{Y}_{\text{LB}}\right|$ \\
    $N_{\text{TR}}$ & the number of elements of $S^Z_{\text{TR}}$ where
    $N_{\text{TR}} = \left|S^{X}_{\text{TR}}\right| =
    \left|S^{Y}_{\text{TR}}\right|$ \\
    $N_{\text{BC}}$ & the number of elements of $S^Z_{\text{BC}}$ where
    $N_{\text{BC}} = \left|S^{X}_{\text{BC}}\right| =
    \left|S^{Y}_{\text{BC}}\right|$ \\
    $N_{\text{UL}}$ & the number of elements of $S^Z_{\text{UL}}$ where
    $N_{\text{UL}} = \left|S^{X}_{\text{UL}}\right| =
    \left|S^{Y}_{\text{UL}}\right|$ \\
    $f(X)$ & the model trained with $S^{X}_{\text{TR}}$ and $S^{Y}_{\text{TR}}$
    \\
    $Y^P$ & the machine learning estimation on $Y$ \\
    \hline
    \hline
    \end{tabular}
    \end{adjustbox}
    \caption{Notation and convention used in this paper for the explanation of our work.}
    \label{tab:notation-1}
    \vspace{-0.5em}
\end{table}
Our notation and convention used in this work are summarized in
Table~\ref{tab:notation-1}.
For the machine learning (ML) estimation, we perform the following two steps:
\begin{enumerate}
\item We train a model $f(X)$ using the data in the $S^X_{\text{TR}}$ and
$S^Y_{\text{TR}}$.
\item We input $X \in S^X_{\text{UL}}$ into the model $f(X)$ and obtain ML
estimation $f(X) = Y^P \approx Y$.
\end{enumerate}
Note that only $Z \in S^Z_{\text{LB}} \subset S^Z$ ($Z=X,Y$) can be used for the
model training, which means that there may exist any kind of unwanted
fluctuations or weird bias due to the \emph{partial} data usage, in principle.
Therefore, using the methodology given in Ref.~\cite{Yoon:2018krb}, we do not
use all $S^Z_{\text{LB}}$ for the model training but use only $S^Z_{\text{TR}}
\subset S^Z_{\text{LB}}$, the training set.
We use remaining $S^Z_{\text{BC}} = S^Z_{\text{LB}} 
\setminus S^Z_{\text{TR}}$ for the
bias correction.
Then we calculate following two estimations:
\begin{align}
    \bar{Y}_{\mathcal{P}1} & = \frac{1}{N_{\textrm{UL}}} \sum_{Y_i \in
    S^Y_{\text{UL}}} Y_i^{P} + \frac{1}{N_{\textrm{BC}}} \sum_{Y_j \in
    S^Y_{\text{BC}}} \left( Y_j - Y_j^{P}\right) \,, \label{eq:P1-1} \\   
    \bar{Y}_{\mathcal{P}2} & = \frac{N_{\textrm{UL}}}{N} \,
    \bar{Y}_{\mathcal{P}1} + \frac{N_{\textrm{LB}}}{N} \, \bar{Y}_{(\text{LB})}
    \quad \text{where} \quad N = N_{\text{LB}} + N_{\text{UL}}\,, \quad
    \bar{Y}_{(\text{LB})} = \frac{1}{N_{\text{LB}}} \sum_{Y_k \in
    S^Y_{\text{LB}}} Y_k \,. \label{eq:P2-1}
\end{align}
Here, $\bar{Y}_{\mathcal{P}1}$ in Eq.~\eqref{eq:P1-1} is the statistical average
on $Y^P$ with the bias correction.
To improve the statistical precision, we further introduce
$\bar{Y}_{\mathcal{P}2}$ in Eq.~\eqref{eq:P2-1}, the weighted average of
$\bar{Y}_{\mathcal{P}1}$ and the original CG results in the labeled set,
$\bar{Y}_{(\text{LB})}$.
To obtain statistical errors, we use the bootstrap resampling method with
$N_{\text{BS}} = 10,000$ where $N_{\text{BS}}$ represents the number of
bootstrap resamples.

\begin{figure}[tb]
    \vspace{-1.5em}
    \begin{adjustbox}{max width=\textwidth}
        \subfigure[$\kappa = 0.13575$, \texttt{ID-0} (the heaviest quark)]{
        \includegraphics[width=0.47\linewidth]{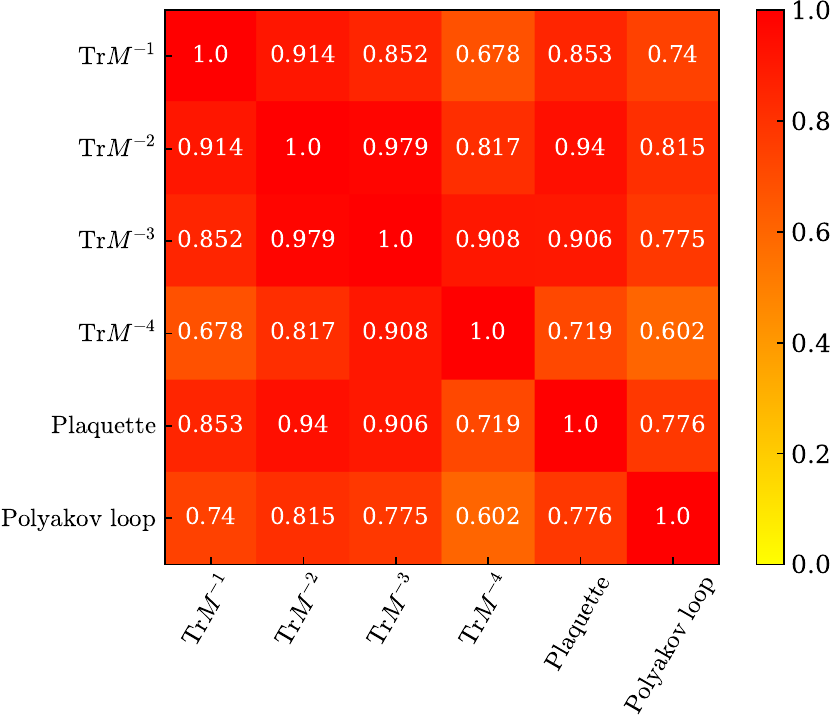}
        \label{fig:subfig-corr-ID-0}
    }
    \hfill
    \subfigure[$\kappa = 0.13585$, \texttt{ID-2} (the lightest quark)]{
        \includegraphics[width=0.47\linewidth]{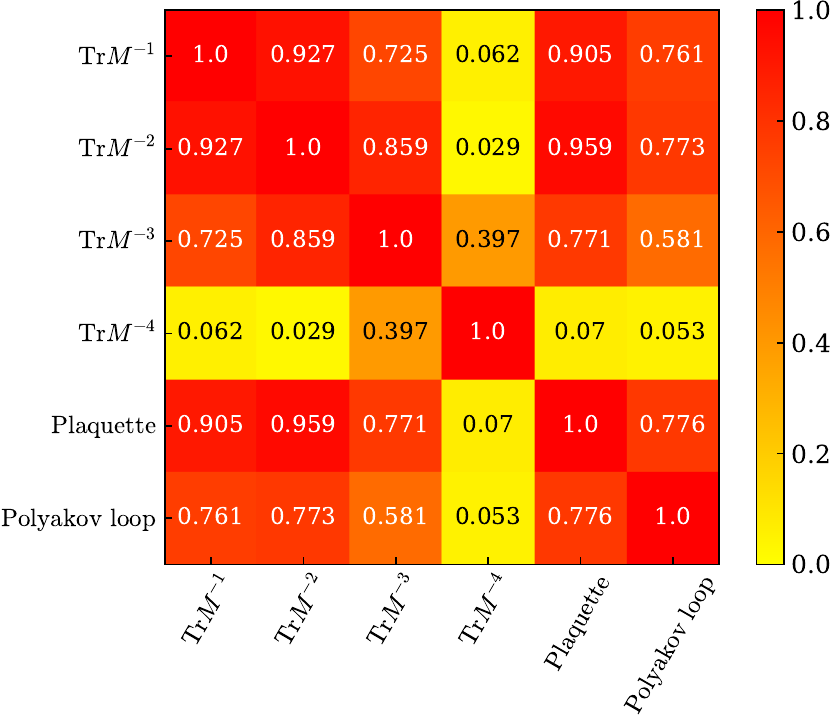}
        \label{fig:subfig-corr-ID-2}
    }
    \end{adjustbox}
    \caption{ Correlation between physical observables.}
    \label{fig:corr-obs-1}
  \end{figure}

For the ML estimation on $Y$ using $X$, the strong correlation
between $X$ and $Y$ is often required.
In the Fig.~\ref{fig:corr-obs-1}, we report the correlation between $\Tr \,
M^{-n}$ ($n=1,2,3,4$), plaquette and Polyakov loop from two datasets:
\texttt{ID-0} (the heaviest quark) and \texttt{ID-2} (the lightest quark).
The red (yellow) color represents strong (weak) correlation between observables.
Note that we need to be careful with $Y = \Tr \, M^{-4}$ case, due to the weak
correlation with other observables such as $X = \Tr \, M^{-m}$ ($m=1,2,3$),
$X=\text{Plaquette}$ and $X=\text{Polyakov loop}$.
Except for it, we observe strong correlations in most cases.
Hence, using the overall tendency of strong correlation, we perform the ML
estimation.
We also check the results for $Y = \Tr \, M^{-4}$ and compare it with the other
analysis results for $Y = \Tr \, M^{-n}$ ($n=1,2,3$).

For the economic efficiency of the ML estimation, that is, to reduce the
computational cost from the linear CG solver, we should use the least possible
$S^Z_{\text{LB}}$ ($Z=X,Y$).
Here, note that we need to use $Y$ as well as $Y^P$ for the bias correction.
This means that we should grant sufficient statistics to the $S^Z_{\text{BC}}$
among the $S^Z_{\text{LB}}$.
In summary, we monitor the following two factors in this work:
\begin{enumerate}
\item We find out minimal $\displaystyle{\mathcal{R}_{\text{LB}} \equiv
\frac{N_{\text{LB}}}{N}}$ where $N = N_{\text{LB}} + N_{\text{UL}}$ and
$\mathcal{R}_{\text{LB}}=5,\,10,\,\cdots,\,50\%$.
\item We find out maximal $\displaystyle{\mathcal{R}_{\text{TR}} \equiv
\frac{N_{\text{TR}}}{N_{\text{LB}}}}$ where $N_{\text{LB}} = N_{\text{TR}} +
N_{\text{BC}}$ and $\mathcal{R}_{\text{TR}}=10,\,20,\,\cdots,\,90\%$.
\end{enumerate}
Here, we also monitor $\mathcal{R}_{\text{TR}} = 0 \%$ ($S_{\text{TR}}^Z =
\emptyset$ for $Z=X,Y$) to check that $S^Y_{\text{LB}}$ itself approaches to the
true answer along the increase of $\mathcal{R}_{\text{LB}}$, that is, we do not
perform ML estimation at $\mathcal{R}_{\text{TR}} = 0 \%$ but observe only the
statistical average and error of the $S^Y_{\text{LB}}$.
We also monitor $\mathcal{R}_{\text{TR}} = 100 \%$ ($S_{\text{BC}}^Z =
\emptyset$) to check what happens when we do not use the bias correction method
in the ML estimation.

For the ML estimation of $Y = \Tr \, M^{-n}$ ($n=1,2,3,4$), we
use the gradient boosting decision tree regression method
\cite{10.1214/aos/1013203451}.
To be specific, we use \texttt{LightGBM} \cite{NIPS2017_6449f44a} framework via
\texttt{JuliaAI/MLJ.jl} \cite{Blaom2020}.
We use 40 boosting stages of depth-3 trees with learning rate of 0.1 and
sub-sampling of 0.7.

\begin{figure}[b!]
    \begin{adjustbox}{max width=\textwidth}
    \subfigure[Score 2]{
        \includegraphics[width=0.3\linewidth]{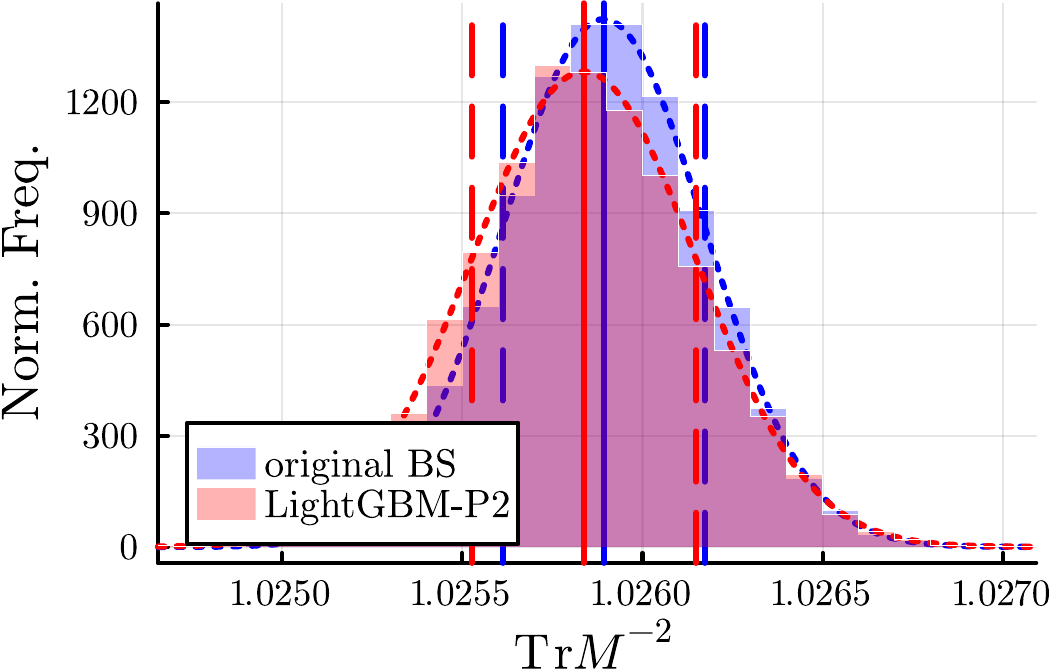}
        \label{subfig:score-2}
    }
    \hfill
    \subfigure[Score 1]{
        \includegraphics[width=0.3\linewidth]{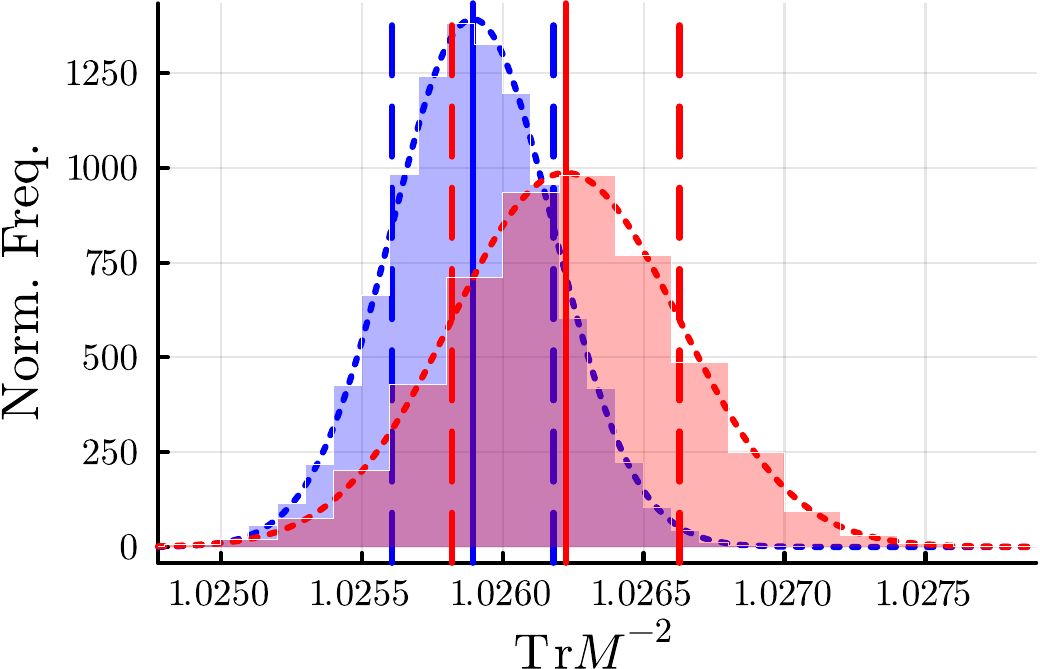}
        \label{subfig:score-1}
    }
    \hfill
    \subfigure[Score 0]{
        \includegraphics[width=0.3\linewidth]{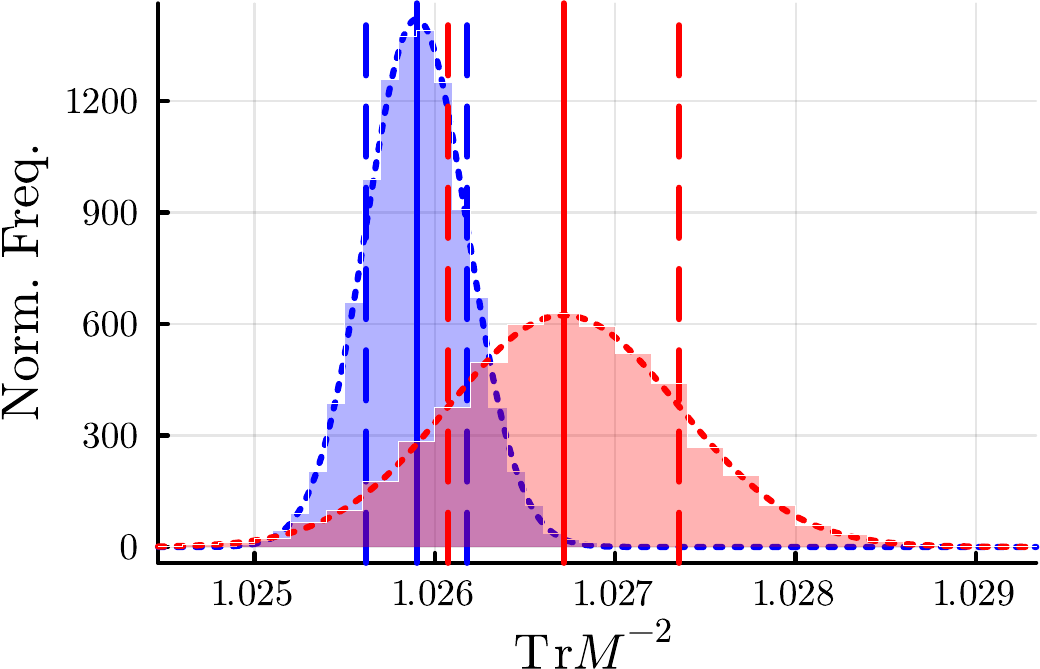}
        \label{subfig:score-0}
    } 
    \end{adjustbox}
    \caption{ Histograms on the three possible cases can be occurred where \blue{blue}
    (\red{red}) histogram represents \blue{original CG result} (\red{ML estimation}).
    Here, data for $X = \Tr \, M^{-1}$ and $Y = \Tr \, M^{-2}$ are used.
    Vertical solid (dotted) lines represent statistical averages ($1\,\sigma$
    errors).}
    \label{fig:eval-1}    
  \end{figure}
To check the usefulness of this method, we compare the ML estimation,
$\{\bar{Y}_\mathcal{Q}, \sigma_\mathcal{Q}\}$
$(\mathcal{Q}=\mathcal{P}_1,\mathcal{P}_2$), with the original CG result,
$\{\bar{Y}_\mathrm{Orig}, \sigma_\mathrm{Orig}\}$.
We prepare following two evaluation criteria (\textbf{EC-}$\mathbf{x}$).
\begin{enumerate}[ label=\textbf{EC-\arabic*}, ref=\textbf{EC-\arabic*},
    itemsep=0.5pt]
    \item \label{it:eval-1} We check whether statistical average of ML
    estimation and original CG results are close to each other or not.
    \begin{enumerate}
        \item If both of $\bar{Y}_{\text{Orig}}$ and
        $\bar{Y}_{\mathcal{Q}}$ agree with $1\,\sigma$ level each
        other, then we grant score 2.
        (Fig.~\ref{fig:eval-1}\subref{subfig:score-2})
        \item If only one of them is included in the other's $1\,\sigma$ error,
        then we grant score 1. (Fig.~\ref{fig:eval-1}\subref{subfig:score-1})
        \item If $\bar{Y}_{\text{Orig}}$ and $\bar{Y}_{\mathcal{Q}}$ do not
        agree with $1\,\sigma$ level each other, then we grant score 0.
        (Fig.~\ref{fig:eval-1}\subref{subfig:score-0})
    \end{enumerate}
    \item \label{it:eval-2} We check whether 
    \begin{align}
        \mathcal{R}_{\sigma} & \equiv \frac{\sigma_{\mathcal{Q}}}{\sigma_{\text{Orig}}}
        \approx 1 \,,
        \label{eq:R-sigma-1}
    \end{align}
    that is, the statistical error of ML estimation, $\sigma_\mathcal{Q}$, is
    close to that of original CG result, $\sigma_\mathrm{Orig}$, or not.
    Currently, we are finding unambiguous explicit criterion for this.
    In this paper, we use $\mathcal{R}_{\sigma} \le 1.2$ which is tentatively
    determined empirically, monitoring our preliminary results. 
\end{enumerate}
Therefore, if an ML estimation got score 2 at the \ref{it:eval-1} and turned out
that $\mathcal{R}_{\sigma} \approx 1$ (tentatively $\mathcal{R}_{\sigma} \le
1.2$ in this paper) at the \ref{it:eval-2}, then this ML estimation can be
thought that it imitates its original CG result as well as possible.

%--------------------------------------------
\section{Results}
\label{sec:res}
%--------------------------------------------

\begin{figure}[tb]
    \begin{adjustbox}{max width=\textwidth}
    \subfigure[Result of \ref{it:eval-1}]{
        \includegraphics[width=0.473\linewidth]{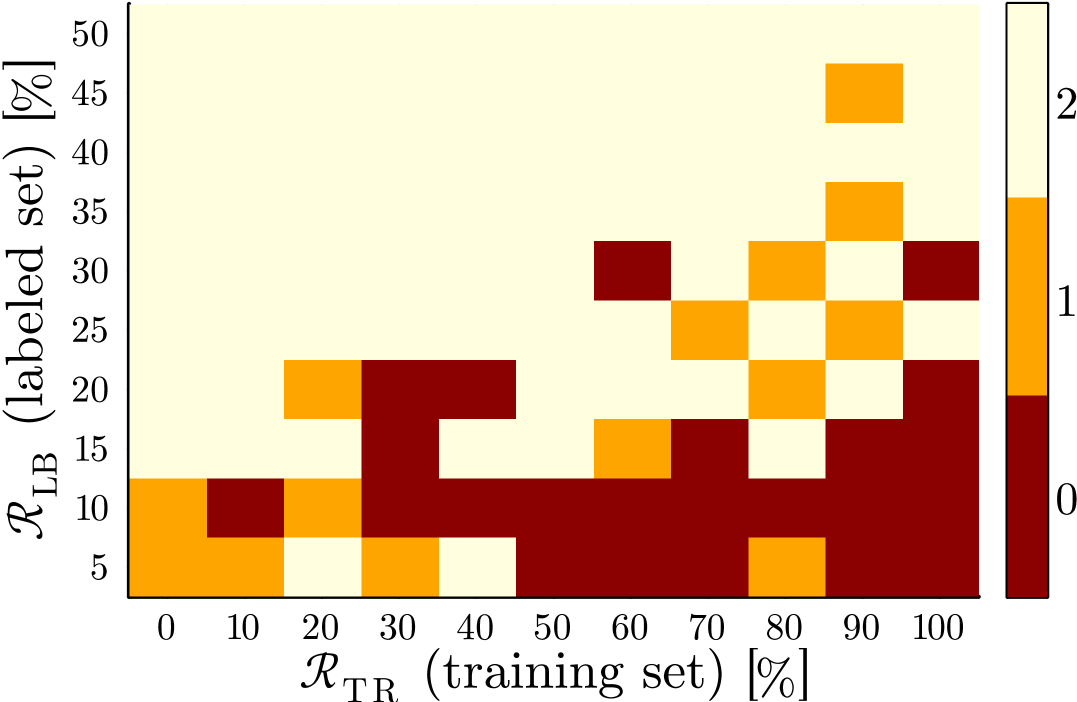}
        \label{fig:subfig-P2-plaq-tr3-ID0-eval1}
    }
    \hfill
    \subfigure[Result of \ref{it:eval-2}]{
        \includegraphics[width=0.473\linewidth]{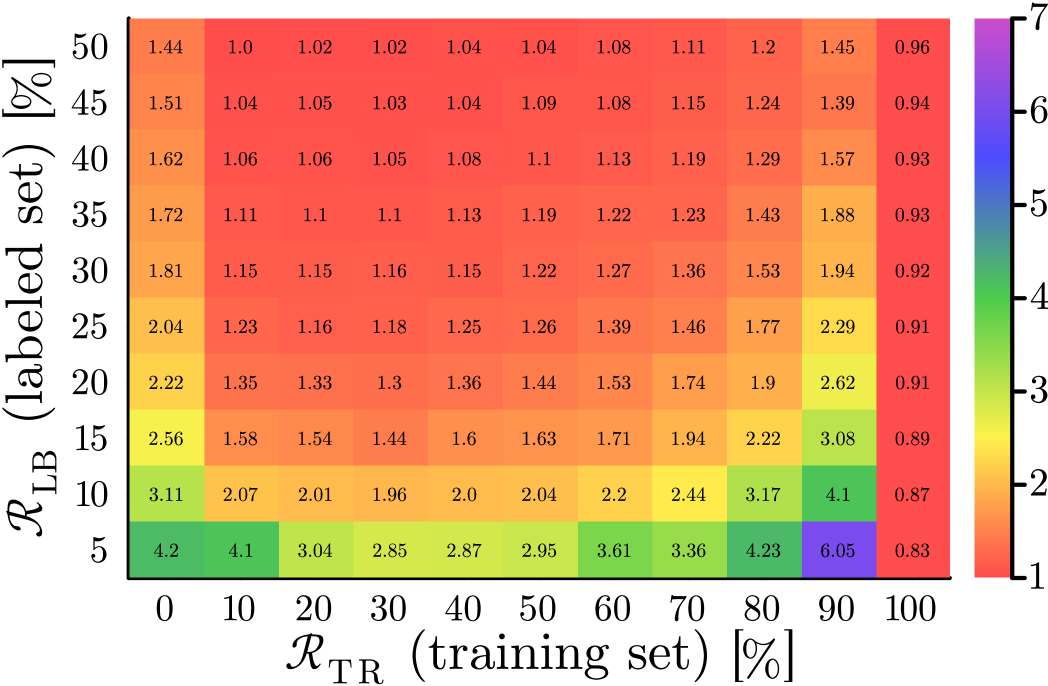}
        \label{fig:subfig-P2-plaq-tr3-ID0-eval2}
    } 
    \end{adjustbox}
    \caption{Results on \subref{fig:subfig-P2-plaq-tr3-ID0-eval1}
    \ref{it:eval-1} and \subref{fig:subfig-P2-plaq-tr3-ID0-eval2}
    \ref{it:eval-2} of $\mathcal{P}2$ estimation using $(X\,,\,Y) =
    (\text{Plaquette} \,,\, \Tr \, M^{-3})$ for \texttt{ID-0} dataset where the
    heaviest quark is used.
    White, orange, red color of \subref{fig:subfig-P2-plaq-tr3-ID0-eval1}
    represents score 2, 1, 0, respectively.
    Here, the magnitude of $\mathcal{R}_{\sigma}$ (Eq.~\eqref{eq:R-sigma-1}) in
    \subref{fig:subfig-P2-plaq-tr3-ID0-eval2} is represented using the color of
    rainbow.
    For example, if $\mathcal{R}_{\sigma} \approx 7$ for certain point, then we
    observe purple-like color there.}
    \label{fig:P2-plaq-tr3-ID0}
  \end{figure}

\begin{figure}[tb]
    \begin{adjustbox}{max width=\textwidth}
    \subfigure[Result of \ref{it:eval-1}]{
        \includegraphics[width=0.473\linewidth]{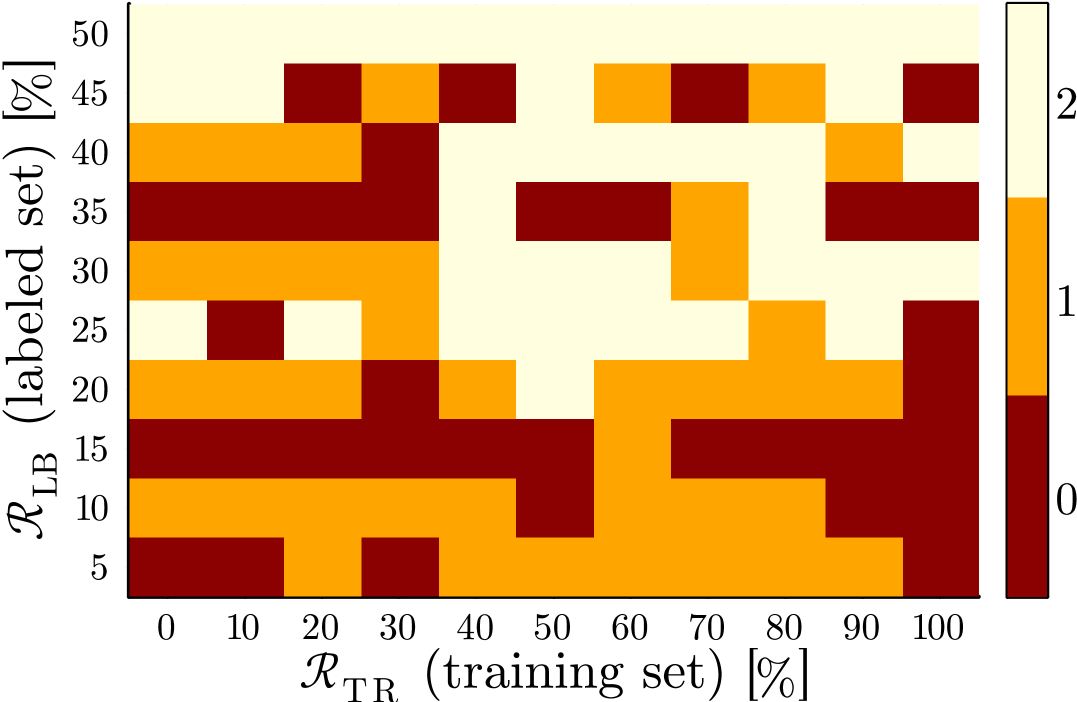}
        \label{fig:subfig-P2-plaq-tr3-ID2-eval1}
    }
    \hfill
    \subfigure[Result of \ref{it:eval-2}]{
        \includegraphics[width=0.473\linewidth]{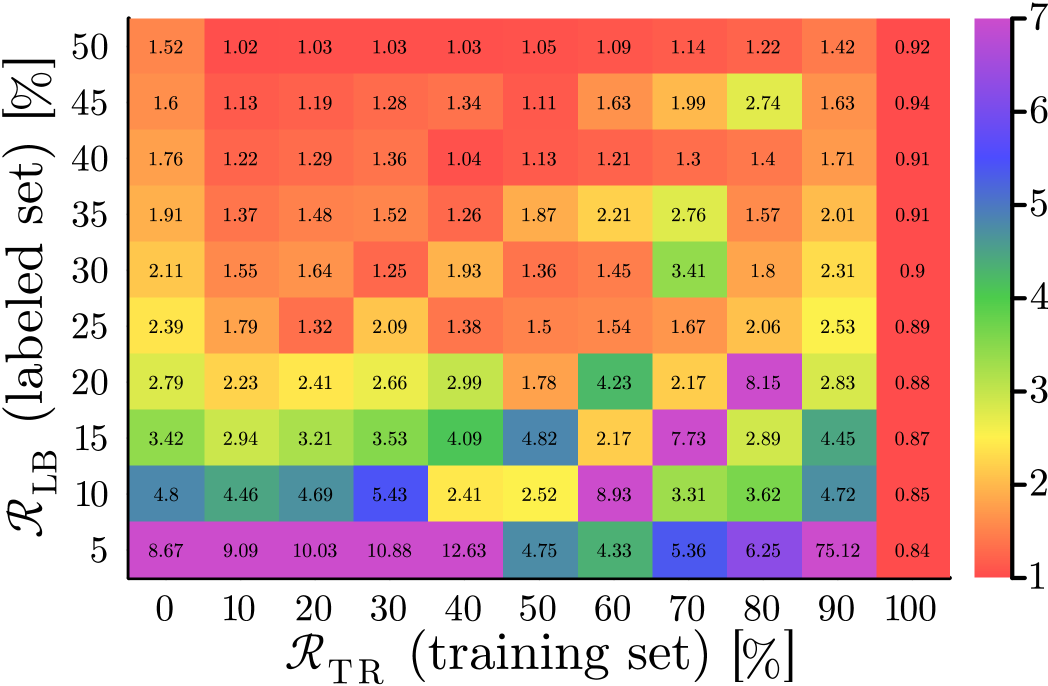}
        \label{fig:subfig-P2-plaq-tr3-ID2-eval2}
    }      
    \end{adjustbox}
    \caption{Results on \subref{fig:subfig-P2-plaq-tr3-ID2-eval1}
    \ref{it:eval-1} and \subref{fig:subfig-P2-plaq-tr3-ID2-eval2}
    \ref{it:eval-2} of $\mathcal{P}2$ estimation using $(X\,,\,Y) = (\text{Plaquette} \,,\, \Tr \, M^{-3})$ for \texttt{ID-2} dataset where the lightest quark is
    used.
    We use same notation as in Fig.~\ref{fig:P2-plaq-tr3-ID0}.}
    \label{fig:P2-plaq-tr3-ID2}
  \end{figure}

\begin{figure}[tb]
    \begin{adjustbox}{max width=\textwidth}
    \subfigure[Result of \ref{it:eval-1}]{
        \includegraphics[width=0.473\linewidth]{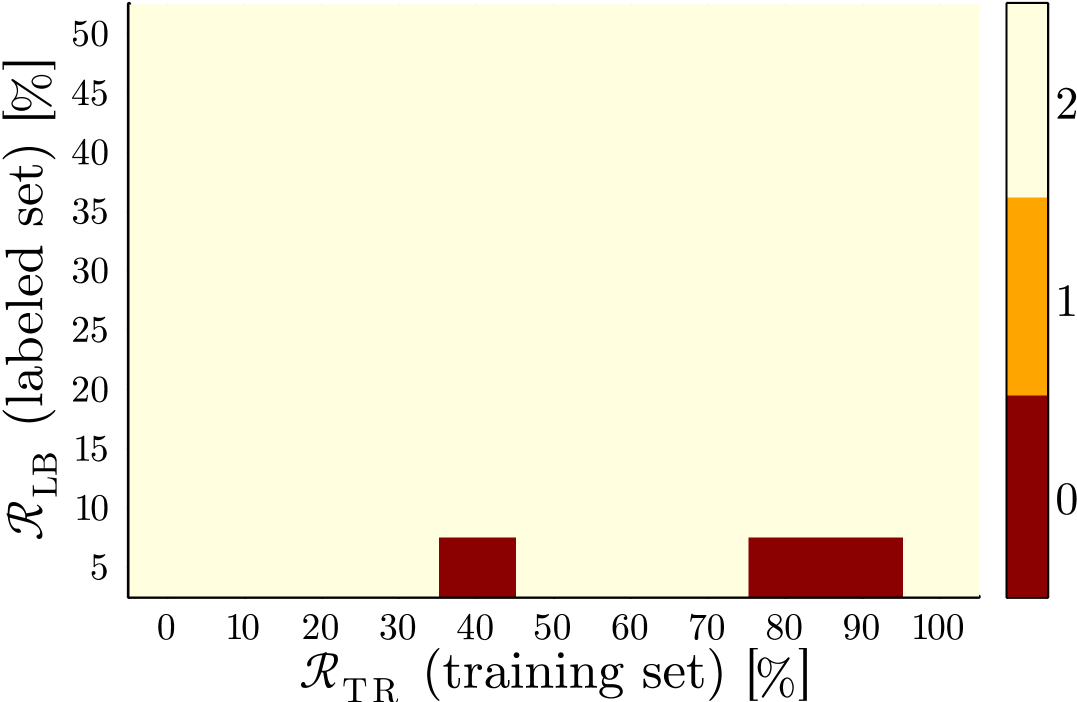}
        \label{fig:subfig-P2-plaq-tr3-ID1-eval1}
    }
    \hfill
    \subfigure[Result of \ref{it:eval-2}]{
        \includegraphics[width=0.473\linewidth]{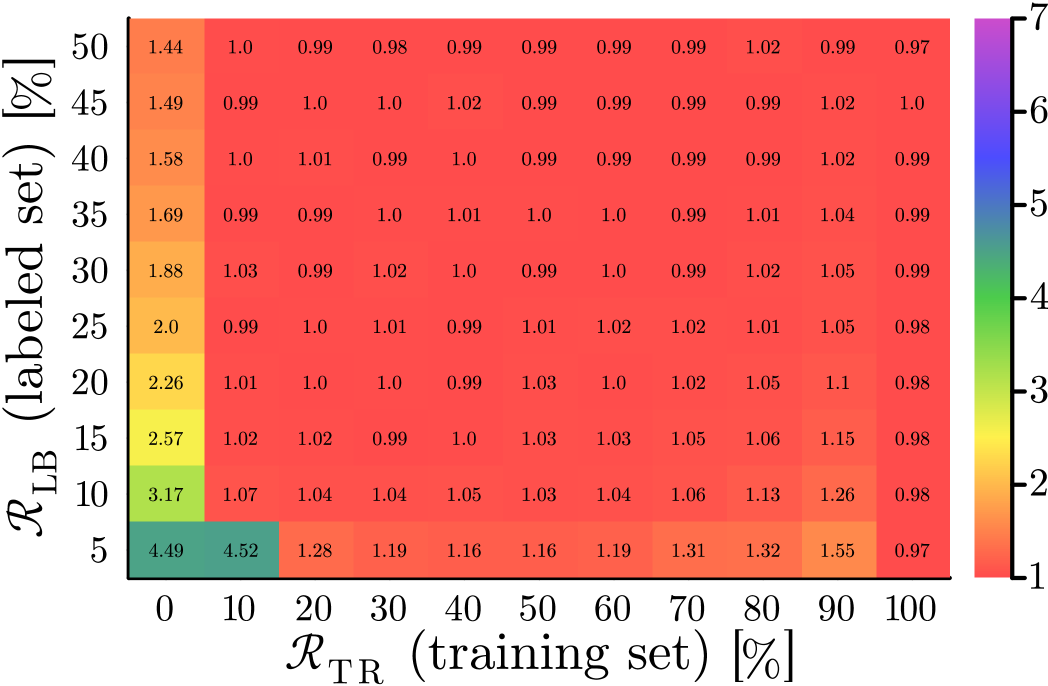}
        \label{fig:subfig-P2-plaq-tr3-ID1-eval2}
    }      
    \end{adjustbox}
    \caption{Results on \subref{fig:subfig-P2-plaq-tr3-ID2-eval1}
    \ref{it:eval-1} and \subref{fig:subfig-P2-plaq-tr3-ID2-eval2}
    \ref{it:eval-2} of $\mathcal{P}2$ estimation using $(X\,,\,Y) = (\text{Plaquette} \,,\, \Tr \, M^{-3})$ for \texttt{ID-1} dataset where the first order phase
    transition observed.
    We use same notation as in Fig.~\ref{fig:P2-plaq-tr3-ID0}.}
    \label{fig:P2-plaq-tr3-ID1}
  \end{figure}

\begin{table}[tb]
    \renewcommand{\arraystretch}{1.2}
    \centering
    \begin{adjustbox}{max width=\textwidth}
      \begin{tabular}{@{\quad}c@{\quad}|@{\quad}l@{\qquad}l@{\quad}l@{\qquad}l@{\qquad}l@{\quad}}
    \hline
    \hline
    $\left\{ \bm{\mathcal{R}_{\text{LB}}}^{\gtrsim}, \mathcal{R}_{\text{TR}}^{\lesssim} \right\}$ 
                       & Plaquette                      & Polyakov loop                  & $\text{Tr} \, M^{-1}$             & $\text{Tr} \, M^{-2}$             & $\text{Tr} \, M^{-3}$ \\
    \hline
    $\text{Tr} \, M^{-1}$ & $\{ {\mathbf{40}} ,\, {50} \}$ & $\{ {\mathbf{40}} ,\, {40} \}$ &                                &                                &   \\ 
    $\text{Tr} \, M^{-2}$ & $\{ {\mathbf{30}} ,\, {40} \}$ & $\{ {\mathbf{35}} ,\, {30} \}$ & $\{ {\mathbf{30}} ,\, {40} \}$ &                                &   \\
    $\text{Tr} \, M^{-3}$ & $\{ {\mathbf{30}} ,\, {40} \}$ & $\{ {\mathbf{40}} ,\, {40} \}$ & $\{ {\mathbf{35}} ,\, {40} \}$ & $\{ {\mathbf{25}} ,\, {70} \}$ &   \\
    $\text{Tr} \, M^{-4}$ & $\{ {\mathbf{45}} ,\, {40} \}$ & $\{ {\mathbf{45}} ,\, {40} \}$ & $\{ {\mathbf{45}} ,\, {40} \}$ & $\{ {\mathbf{35}} ,\, {40} \}$ & $\{ {\mathbf{25}} ,\, {40} \}$ \\
    \hline
    \hline
    \end{tabular}
    \end{adjustbox}
    \caption{Results on $\left\{ \bm{\mathcal{R}_{\text{LB}}}^{\gtrsim}, \mathcal{R}_{\text{TR}}^{\lesssim} \right\}$ of $\mathcal{P}2$ estimation for \texttt{ID-0} dataset where the heaviest quark is used. 
    Here, the column and row represent the $X$ (input) and $Y$ (output),
    respectively.   
    For example, $\left\{ \bm{\mathcal{R}_{\text{LB}}}^{\gtrsim},
    \mathcal{R}_{\text{TR}}^{\lesssim} \right\} = \left\{ \mathbf{30}, 40
    \right\}$ represents that ML estimation got consistent score 2 at
    \ref{it:eval-1} and showed $\mathcal{R}_{\sigma} \le 1.2$ at \ref{it:eval-2}
    in the $\mathcal{R}_{\text{LB}} \gtrsim 30 \%$ and $\mathcal{R}_{\text{TR}}
    \lesssim 40 \%$ region.}
    \label{tab:ID-0-res}
\end{table} % \diagbox{.}{.} \diagbox[dir=NE]{}{}
Here we show our preliminary results on the ML estimation of $Y = \mathrm{Tr} \,
M^{-n}$.
Note that we use single $X$ for the ML estimation of $Y$ in this analysis.
As examples, we show results on \ref{it:eval-1} and \ref{it:eval-2} using
$(X\,,\,Y) = (\text{Plaquette} \,,\, \Tr \, M^{-3})$ for \texttt{ID-0}
dataset (Fig.~\ref{fig:P2-plaq-tr3-ID0}), \texttt{ID-1} dataset
(Fig.~\ref{fig:P2-plaq-tr3-ID1}), and \texttt{ID-2} dataset
(Fig.~\ref{fig:P2-plaq-tr3-ID2}).

We report our results on \texttt{ID-0} dataset where the heaviest quark is used
in the measurement.
In Fig.~\ref{fig:P2-plaq-tr3-ID0}\subref{fig:subfig-P2-plaq-tr3-ID0-eval1}, we
observe that we obtain score 2 consistently in the region of
$\mathcal{R}_{\text{LB}} \gtrsim 30 \%$ and $\mathcal{R}_{\text{TR}} \lesssim 50
\%$ (\ref{it:eval-1}).
In Fig.~\ref{fig:P2-plaq-tr3-ID0}\subref{fig:subfig-P2-plaq-tr3-ID0-eval2}, we
observe that we obtain $\mathcal{R}_{\sigma} \le 1.2$ in the region of
$\mathcal{R}_{\text{LB}} \gtrsim 30 \%$ and $\mathcal{R}_{\text{TR}} \lesssim 40
\%$ (\ref{it:eval-2}).
As we can see from $\mathcal{R}_{\text{TR}} = 100 \%$ column ($S_{\text{BC}}^Z =
\emptyset$ for $Z=X,Y$) of
Fig.~\ref{fig:P2-plaq-tr3-ID0}\subref{fig:subfig-P2-plaq-tr3-ID0-eval1}, we need
$\mathcal{R}_{\text{LB}} \gtrsim 35 \%$, slightly larger
$\mathcal{R}_{\text{LB}}$ than those of $S_{\text{BC}}^Z \neq \emptyset$ case.

In Table \ref{tab:ID-0-res}, we report our analysis on all possible
cases with single input ($X$) and output ($Y$) for \texttt{ID-0} dataset.
We found that the ML estimation on $Y=\Tr \, M^{-n}$ works well for
$\mathcal{R}_{\text{LB}} \gtrsim 40 \%$ even when we use
$X=\text{Plaquette}$ and $X=\text{Polyakov loop}$. 
We also found that the ML estimation on $Y=\Tr \, M^{-4}$ needs more ratio of
labeled set ($\mathcal{R}_{\text{LB}}$) than $Y=\Tr \, M^{-n}$ ($n=1,2,3$).
However, we need $\mathcal{R}_{\text{LB}} \gtrsim 25 \%$ when $(X\,,\,Y)=(\Tr \,
M^{-2}\,,\,\Tr \, M^{-3})$ and $(X\,,\,Y)=(\Tr \, M^{-3}\,,\,\Tr \, M^{-4})$.

\begin{table}[tb]
    \renewcommand{\arraystretch}{1.2}
    \centering
    \begin{adjustbox}{max width=\textwidth}
      \begin{tabular}{@{\quad}c@{\quad}|@{\quad}l@{\qquad}l@{\quad}l@{\qquad}l@{\qquad}l@{\quad}}
    \hline
    \hline
    $\left\{ \bm{\mathcal{R}_{\text{LB}}}^{\gtrsim}, \mathcal{R}_{\text{TR}}^{\lesssim} \right\}$
                       & Plaquette                      & Polyakov loop                  & $\text{Tr} \, M^{-1}$             & $\text{Tr} \, M^{-2}$ & $\text{Tr} \, M^{-3}$ \\
    \hline
    $\text{Tr} \, M^{-1}$ & $\{ {\mathbf{30}} ,\, {40} \}$ & $\{ {\mathbf{35}} ,\, {30} \}$ &                                &                    &      \\
    $\text{Tr} \, M^{-2}$ & $\{ {\mathbf{40}} ,\, {80} \}$ & $\{ {\mathbf{40}} ,\, {40} \}$ & $\{ {\mathbf{30}} ,\, {50} \}$ &                    &      \\
    $\text{Tr} \, M^{-3}$ & N.A.                           & N.A.                           & N.A.                           & N.A.               &      \\
    $\text{Tr} \, M^{-4}$ & N.A.                           & N.A.                           & N.A.                           & N.A.               & N.A. \\
    \hline
    \hline
    \end{tabular}
  \end{adjustbox}
  \caption{Results on $\left\{ \bm{\mathcal{R}_{\text{LB}}}^{\gtrsim}, \mathcal{R}_{\text{TR}}^{\lesssim} \right\}$ of $\mathcal{P}2$ estimation for \texttt{ID-2} dataset where the lightest quark is used.
    We use same notation as in Table \ref{tab:ID-0-res}.
    Here, ``N.A.'' means that we cannot find consistent good region for \ref{it:eval-1} and \ref{it:eval-2}.}
      \label{tab:ID-2-res}
\end{table} % \diagbox{.}{.} \diagbox[dir=NE]{}{}
Next, we report our results on \texttt{ID-2} dataset where the lightest
quark is used in the measurement.
We cannot find consistent score-2 region (\ref{it:eval-1})
from Fig.~\ref{fig:P2-plaq-tr3-ID2}\subref{fig:subfig-P2-plaq-tr3-ID2-eval1}.
We also cannot find consistent $\mathcal{R}_{\sigma} \le 1.2$ region
(\ref{it:eval-2}) from
Fig.~\ref{fig:P2-plaq-tr3-ID2}\subref{fig:subfig-P2-plaq-tr3-ID2-eval2}.
The ML estimation on $Y = \Tr \, M^{-3}$ using $X = \text{Plaquette}$ does not
work well for \texttt{ID-2} dataset.

In Table \ref{tab:ID-2-res}, we report our analysis on all possible
cases with single input ($X$) and output ($Y$) for \texttt{ID-2} dataset.
We found that only the ML estimation on $Y=\Tr \, M^{-n}$ ($n=1,2$) works well
for $\mathcal{R}_{\text{LB}} \gtrsim 40 \%$.
The results on \ref{it:eval-1} and \ref{it:eval-2} for $Y=\Tr \, M^{-n}$
($n=3,4$) are all similar with Fig.~\ref{fig:P2-plaq-tr3-ID2}.

\begin{table}[tb]
    \renewcommand{\arraystretch}{1.2}
    \centering
    \begin{adjustbox}{max width=\textwidth}
      \begin{tabular}{@{\quad}c@{\quad}|@{\quad}l@{\qquad}l@{\quad}l@{\qquad}l@{\qquad}l@{\quad}}
    \hline
    \hline
    $\left\{ \bm{\mathcal{R}_{\text{LB}}}^{\gtrsim}, \mathcal{R}_{\text{TR}}^{\lesssim} \right\}$ 
                       & Plaquette                      & Polyakov loop                  & $\text{Tr} \, M^{-1}$             & $\text{Tr} \, M^{-2}$             & $\text{Tr} \, M^{-3}$ \\
    \hline
    $\text{Tr} \, M^{-1}$ & $\{ {\mathbf{10}} ,\, {80} \}$ & $\{ {\mathbf{10}} ,\, {70} \}$ &                                &                                &   \\
    $\text{Tr} \, M^{-2}$ & $\{ {\mathbf{10}} ,\, {90} \}$ & $\{ {\mathbf{10}} ,\, {70} \}$ & $\{ {\mathbf{10}} ,\, {80} \}$ &                                &   \\
    $\text{Tr} \, M^{-3}$ & $\{ {\mathbf{10}} ,\, {80} \}$ & $\{ {\mathbf{10}} ,\, {70} \}$ & $\{ {\mathbf{10}} ,\, {70} \}$ & $\{ {\mathbf{10}} ,\, {90} \}$ &   \\
    $\text{Tr} \, M^{-4}$ & $\{ {\mathbf{40}} ,\, {70} \}$ & $\{ {\mathbf{40}} ,\, {80} \}$ & $\{ {\mathbf{40}} ,\, {80} \}$ & $\{ {\mathbf{40}} ,\, {80} \}$ & $\{ {\mathbf{40}} ,\, {80} \}$ \\
    \hline
    \hline
    \end{tabular}
    \end{adjustbox}
    \caption{Results on $\left\{ \bm{\mathcal{R}_{\text{LB}}}^{\gtrsim}, \mathcal{R}_{\text{TR}}^{\lesssim} \right\}$ of $\mathcal{P}2$ estimation for \texttt{ID-1} dataset where the first order phase transition is observed.
    We use same notation as in Table \ref{tab:ID-0-res}.}
    \label{tab:ID-1-res}
\end{table} % \diagbox{.}{.} \diagbox[dir=NE]{}{}
Finally, we report our results on \texttt{ID-1} dataset where the first order
phase transition is observed.
In Fig.~\ref{fig:P2-plaq-tr3-ID1}\subref{fig:subfig-P2-plaq-tr3-ID1-eval1}, we
observe that we obtain score 2 consistently in the region of
$\mathcal{R}_{\text{LB}} \gtrsim 10 \%$ and $\mathcal{R}_{\text{TR}} \lesssim 90
\%$ (\ref{it:eval-1}).
In Fig.~\ref{fig:P2-plaq-tr3-ID1}\subref{fig:subfig-P2-plaq-tr3-ID1-eval2}, we
observe that we obtain $\mathcal{R}_{\sigma} \le 1.2$ in the region of
$\mathcal{R}_{\text{LB}} \gtrsim 10 \%$ and $\mathcal{R}_{\text{TR}} \lesssim 80
\%$ (\ref{it:eval-2}).

In Table \ref{tab:ID-1-res}, we report our analysis on all possible cases with
single input ($X$) and output ($Y$) for \texttt{ID-1} dataset.
We found that the ML estimation on $Y=\Tr \, M^{-n}$
($n=1,2,3$) works well for $\mathcal{R}_{\text{LB}} \gtrsim 10 \%$ even when we
use $X=\text{Plaquette}$ and $X=\text{Polyakov loop}$. 
On the other hand, the ML estimation on $Y=\Tr \, M^{-4}$ needs more ratio of
labeled set ($\mathcal{R}_{\text{LB}}$) than $Y=\Tr \, M^{-n}$ ($n=1,2,3$),
\emph{i.e.}, $\mathcal{R}_{\text{LB}} \gtrsim 40 \%$.

%--------------------------------------------
\section{Summary and to-do list}
\label{sec:todo}
%--------------------------------------------

We performed preliminary analysis on ML estimation on $Y =
\Tr \, M^{-n}$ ($n=1,2,3,4$) using $X = \Tr \, M^{-m}$ ($m < n$), $X =
\text{Plaquette}$ and $X = \text{Polyakov loop}$.
Here, we used the gradient boosting decision tree regression method
\cite{10.1214/aos/1013203451}.

With the heaviest quark (\texttt{ID-0} dataset), we observed that the ML
estimation of $Y = \Tr \, M^{-n}$ ($n=1,2,3$) showed consistently good results
at $\mathcal{R}_{\text{LB}} \gtrsim 40 \%$ where $Y = \Tr \, M^{-4}$ required
slightly more $\mathcal{R}_{\text{LB}}$ than $n=1,2,3$ cases.
With the lightest quark (\texttt{ID-2} dataset), we observed that only the ML
estimation of $Y = \Tr \, M^{-n}$ ($n=1,2$) works well at
$\mathcal{R}_{\text{LB}} \gtrsim 40 \%$.
On the other hand, the \texttt{ID-1} dataset where the first order phase
transition is observed, we observed that the ML estimation of $Y = \Tr \,
M^{-n}$ ($n=1,2,3$) showed consistently good results at $\mathcal{R}_{\text{LB}}
\gtrsim 10 \%$.
However, $Y = \Tr \, M^{-4}$ still required $\mathcal{R}_{\text{LB}} \gtrsim 40
\%$ even in this dataset.

In this preliminary result with three datasets, we observed that ML estimation
of $Y = \Tr \, M^{-n}$ works well with heavier quark mass.
Especially, we observed that the ML estimation of $Y = \Tr \, M^{-n}$ works
quite well when the first order phase transition is observed at the dataset.
However, the quality of $Y = \Tr \, M^{-4}$ estimation is not good comparing
 with $Y = \Tr \, M^{-n}$ ($n=1,2,3$).
To get better $Y = \Tr \, M^{-4}$ estimation, we need to use $X = \Tr \,
M^{-3}$, for example (except for \texttt{ID-2} dataset where the lightest quark
is used).

In this paper, we used single input ($X$) for the ML estimation of $Y = \Tr
\, M^{-n}$ ($n=1,2,3,4$).
We need to check whether we get better results when we use multiple inputs: for
example, we use $X = \{ \text{Plaquette}  \,, \Tr \, M^{-1} \}$ for the ML
estimation of $Y = \Tr \, M^{-4}$.
This work is in progress.

We need to check whether we can obtain reliable cumulants of chiral order
parameters such as susceptibility, skewness, kurtosis \cite{Ohno:2018gcx} using
ML estimation.
This work is also in progress.
% \cite{Kuramashi:2016kpb,
% Jin:2017jjp, Ohno:2018gcx, Nakamura:2019gyy, Kuramashi:2020meg}

%--------------------------------------------
\acknowledgments

The work of A.~T.~was partially supported by JSPS KAKENHI Grant Numbers 20K14479,
22H05111 and 22K03539. 
A.~T.~and H.~O.~were partially supported by JSPS KAKENHI Grant Number 22H05112. 
This work was partially supported by MEXT as ``Program for Promoting Researches
on the Supercomputer Fugaku'' (Grant Number JPMXP1020230411, JPMXP1020230409).
%--------------------------------------------

%--------------------------------------------
%--------------
% bibliography
%--------------
\bibliography{ref}
%--------------------------------------------

%--------------------------------------------
\end{document}